\documentclass[12pt,a4]{article}

\usepackage{amsfonts,amsmath,amssymb,enumerate,epsfig,pict2e}
\usepackage{theorem,cite}
\usepackage{color}
\usepackage{graphicx}

\numberwithin{equation}{section}
\newtheorem{definition}{Definition}[section]

\newtheorem{proposition}[definition]{Proposition}

\newcommand{\prf}{\underline{Proof:}\ }
\newcommand{\finprf}{\null \hfill {\rule{5pt}{5pt}}\\ \null}
\newcommand{\ie}{{\it i.e.}\ }

\newcommand{\be}{\begin{equation}}
\newcommand{\ee}{\end{equation}}
\newcommand{\beq}{\begin{equation}}
\newcommand{\eeq}{\end{equation}}
\newcommand{\beu}{\begin{equation*}}
\newcommand{\eeu}{\end{equation*}}
\newcommand{\bea}{\begin{eqnarray}}
\newcommand{\eea}{\end{eqnarray}}
\newcommand{\beano}{\begin{eqnarray*}}
\newcommand{\eeano}{\end{eqnarray*}}
\newcommand{\beaa}{\begin{eqnarray*}}
\newcommand{\eeaa}{\end{eqnarray*}}
\newcommand{\bmx}{\begin{pmatrix}}
\newcommand{\emx}{\end{pmatrix}}

\newcommand{\ri}{\rm i}
\newcommand{\rd}{{\rm d}}
\def\ft{{\widetilde f}}
\def\e{{\rm e}}

\newcommand{\eps}{\varepsilon}

\def\cA{{\cal A}}         \def\cB{{\cal B}}

\def\fa{{\mathfrak a}}

\def\fs{{\mathfrak s}}

\newcommand{\CC}{{\mathbb C}}

\newcommand{\RR}{\mbox{${\mathbb R}$}}
\newcommand{\bS}{\mbox{${\mathbb S}$}}

\newcommand{\ZZ}{{\mathbb Z}}
\newcommand{\1}{\mbox{\hspace{.0em}1\hspace{-.24em}I}}

\newcommand{\prt}{\partial}

\newcommand{\mb}[1]{\quad\mbox{#1}\quad}

\newcommand{\half}{\frac{1}{2}}

\newcommand{\nonu}{\nonumber\\}

\newcommand{\otot}{(\mbox{\scriptsize{out-out}})}
\newcommand{\otin}{(\mbox{\scriptsize{out-in}})}
\newcommand{\inot}{(\mbox{\scriptsize{in-out}})}
\newcommand{\inin}{(\mbox{\scriptsize{in-in}})}
\newcommand{\sloc}{S}

\def\tmu{{\widetilde \mu}}

 \usepackage{color}

\newcommand{\vertex}[1]{\begin{picture}(30,30)
\put(10,10){\circle*{3}}\put(7,20){#1}
\put(10,10){\line(2,1){30}}\put(10,10){\line(-2,1){30}}\put(10,10){\line(0,-1){30}}
\end{picture}}

\newcommand{\vertexD}[1]{\begin{picture}(30,30)
\put(10,15){\circle*{3}}\put(7,0){#1}
\put(10,15){\line(2,-1){30}}\put(10,15){\line(-2,-1){30}}\put(10,15){\line(0,1){30}}
\end{picture}}

\begin{document}

\renewcommand{\thefootnote}{\arabic{footnote}}
\setcounter{footnote}{0}
\newpage
\setcounter{page}{0}

\markright{\today\dotfill \jobname\dotfill }
 \pagestyle{empty}
\setcounter{page}{0}

\vfill

\begin{center}

{\Large\textbf{Exact scattering matrix of graphs in magnetic field and quantum noise}}

\vspace{1cm}

{\large \textbf{Vincent Caudrelier$^a$, Mihail Mintchev$^b$ and Eric Ragoucy$^c$\footnote{email: v.caudrelier@city.ac.uk, mintchev@df.unipi.it, eric.ragoucy@lapth.cnrs.fr}}}

\vfill

\emph{\large
${}^a$ Department of Mathematical Science, City University London, \\
Northampton Square, London EC1V 0HB, UK \\[1.2ex]
${}^b$ Istituto Nazionale di Fisica Nucleare and Dipartimento di Fisica dell'Universit\`a di Pisa, 
Largo Pontecorvo 3, 56127 Pisa, Italy\\[1.2ex] 
${}^c$ LAPTh, Laboratoire d'Annecy-le-Vieux de Physique Th\'eorique, \\
CNRS, Universit\'e de Savoie,   
 BP 110, \\74941 Annecy-le-Vieux Cedex, France}

\vfill
\vfill

\begin{abstract}
We consider arbitrary quantum wire networks modelled by finite, noncompact, connected quantum graphs 
in the presence of an external magnetic field.
 We find a general formula for the total scattering matrix of the network
in terms of its local scattering properties and its metric structure. This is applied to a quantum ring with $N$ external edges. 
Connecting the external edges of the ring to heat reservoirs, we study the quantum transport on the graph
in ambient magnetic field. We consider two types of dynamics on the ring: the free Schr\"odinger and 
the free massless Dirac equations. For each case, a detailed study of the thermal noise is performed analytically. 
 Interestingly enough, in presence of a magnetic field, the standard linear Johnson-Nyquist law for the low temperature behaviour of the thermal noise becomes nonlinear.
The precise regime of validity of this effect is 
discussed and a typical signature of the underlying dynamics is observed.
\end{abstract}
\end{center}
\vfill

\rightline{IFUP-TH/2013-26\qquad}
\rightline{LAPTH-074/13\qquad}
\rightline{January 2014\qquad}
\rightline{}

\newpage
\pagestyle{plain}

\section{Introduction}

The study of transport properties on networks of quantum wires offers both a basis for deep theoretical insight as well as a benchmark for physical applications 
and comparisons with experiments. The wealth of literature on various aspects of this topic indicates that it is a very lively 
and important subject. Up to now, most of the efforts have been focused of the computation of the one-point current correlator $\langle j\rangle$ in order to extract the conductance properties. Several methods have been used to achieve this: bosonization techniques \cite{KF-a,SS-a,NF-a,SD-a,MW-a, PPIL-a, KD-a,SRS-a,Mintchevetal}, renormalization group analysis \cite{Yi-b,LRS-b,EMAB-b,DRS-b}, conformal field theory \cite{COA-c,Affetal} or Landauer-B\"uttiker formalism \cite{CTE-d}.
In this paper, we study the two-point function $\langle j\,j\rangle$ and extract from it the quantum noise.
Since we are looking for finite size effects, the most suitable formalism is
 the Landauer-B\"uttiker approach \cite{la-57,bu-86}. The latter
 relies crucially on the scattering properties of the system under consideration. 
The typical setup is to connect the sample of interest to a number of heat 
reservoirs which drive the system out of equilibrium. Under certain conditions, they bring it to a steady state where physical quantities such as the conductance, 
the current or the noise can be computed. The key to make predictions is to gain information on the scattering through the sample since all the formulas in the 
Landauer-B\"uttiker approach rely on the transmission amplitudes from one external lead to another via the sample. 

It turns out that in the case where the sample is taken to be a finite, connected quantum graph representing a model for one-dimensional transport in a network 
of quantum wires, one has powerful methods to compute the total scattering matrix of the resulting sample, knowing only the metric structure and the 
local scattering at each node (or vertex) of the graph \cite{CR}. In turn, the local scattering is completely determined by the classification of the self-adjoint 
extensions of the underlying differential operator used to model the dynamics of the particles one is interested in. For instance, in the case of the free Schr\"odinger 
equation, one can use the results of \cite{KS}. This means that the class of total scattering matrices of a given arbitrary quantum graph is completely 
determined. It can then be used to study transport analytically and exactly on such structures. 

The presence of a magnetic field can have dramatic effects on quantum transport properties and it is 
one the purposes of this paper to extend the method of \cite{CR} to the case with external magnetic field. The total scattering matrix (\ref{expression_Stot}) has the same structure as in the case without magnetic field but all the information 
on the magnetic flux is encoded in the matrix describing the propagation of particle modes along the wires (or edges) of the graph. In the case where 
the graph contains loops or cycles, the magnetic field is captured in fluxes through these structures and can strongly influence 
transport as seen for instance in the resonant tunneling effect on a quantum ring \cite{biy-84,CRM}. Equipped with our computationally
efficient formula for the total scattering matrix, one can in principle study the quantum transport properties on an arbitrary graph and for several kinds of 
compatible dynamics on the edges, like the free Schr\"odinger, the free Dirac equation or the Tomonaga-Luttinger model. 
In this paper, we carry out this program from start to end by focusing 
our attention on the following combination of graph/systems/physical quantity of interest: a quantum ring connected to an arbitrary number of reservoirs with the 
free Schr\"odinger or free massless Dirac equation on the edges for which we study the properties of the pure thermal noise. In particular, we show that 
the low temperature behaviour of the pure thermal noise is strongly affected by the presence of a magnetic flux through the ring. The usual linear temperature 
dependence (Johnson-Nyquist law) is dramatically changed to a quadratic (Schr\"odinger) or cubic (Dirac) dependence. These results were presented numerically
in the free Schr\"odinger case on a ring connected to only three reservoirs in a companion letter \cite{CRM}. 
The present paper provides the full analytical treatment in the general case of $N$ reservoirs and for both the Schr\"odinger and Dirac cases. This relies on 
thorough analysis of the total scattering matrix and in particular of what we call its fundamental eigenvalue. 

The paper is organised as follows. In section \ref{general_method}, we present the extension of the method of \cite{CR} to compute the total scattering matrix 
of an arbitrary finite, connected quantum graph in the presence of a magnetic field. This is then applied to the case of a quantum ring with $N$ external edges. In 
this particular case, we show that all the relevant information about the total scattering matrix can be reduced to a single eigenvalue, the fundamental eigenvalue.
In section \ref{phys_mod}, we present the setup whereby the quantum graph is connected to external reservoirs thus allowing us to use the usual Landauer-B\"uttiker 
transport formulas for the current and the noise. Using our detailed knowledge of the fundamental eigenvalue, we are able to study exactly the behaviour 
of the pure thermal noise. This is done both for the Schr\"odinger and Dirac cases. The main result there is the derivation of an important effect of the 
magnetic flux on the power law appearing in the low temperature behaviour of the thermal noise. This provides an analytical confirmation of the similar 
numerical observation made in \cite{CRM}. Section \ref{conclu} contains a discussion of the results, our conclusions and some directions for future investigations.

\section{Total scattering matrix of a graph in magnetic field\label{general_method}}

\subsection{General method\label{sect:general}}

In \cite{CR}, two of the authors presented a general method to compute systematically and efficiently the total scattering matrix
of a quantum graph knowing only the local scattering matrices and the metric structure of the graph, in the absence of 
a magnetic field. From a mathematical point of view, this was achieved by using the algebraic structure of the so-called 
Reflection-Transmission algebras \cite{MRS}. From the physical point of view, it amounts to a systematic generalization of the 
scattering formalism \`a la Landauer-B\"uttiker.

In this paper, we generalise the formalism to incorporate an external magnetic field and we adapt the procedure explained in \cite{KS1}
to the method of \cite{CR}. To make this paper more self-contained, we present the main setting and notations as in Section 2.1 of \cite{CR} that are 
useful in the present paper.

We consider a finite, noncompact, and connected graph with $N$ vertices that we label with 
$a=1,\ldots,N$ and with internal and external edges. The graph is 
compact if it has no external edges. 
At each vertex $a$ are attached possibly
several edges. One can endow the graph with a metric structure: 
the \textit{external edges} are associated to infinite half-lines and are 
connected to a unique vertex; the 
\textit{internal edges} are associated to intervals of finite length and 
connect two vertices, possibly not distinct. In the 
case where an internal edge connects the same vertex, we call it a loop 
(also called tadpole in the literature). Two edges are 
adjacent if they are connected by an internal edge. We consider a connected 
graph \ie a graph such that for any two vertices $a$, 
$b$ there is a sequence $\{a_1=a,a_2,\dots,a_q=b\}$ 
of adjacent vertices. We define an 
orientation on the edges, and in the case of internal edges, 
$(ab)$ will define an edge   
with orientation from vertex $a$ to vertex $b$. By convention, 
external edges $(a 0)$ are always oriented 
from the vertex to infinity. On each of these 
edges, we attach operators or \textit{modes}
$$
\fa_{j}^{ab}(k)\quad j=1,\ldots,N_{ab}\ ;\ 
b=0,1,\ldots,N\ ;\ a=1,\ldots,N\ ; \ a\neq b\,,
$$
$k$ being an orientation dependent parameter which has the interpretation 
of a momentum or a rapidity in applications to quantum fields on graphs and 
with the following conventions: 
\begin{itemize}
\item$a=1,2,\ldots,N$ denotes the vertex to which 
the edge is attached; 
\item$b=0,1,2,\ldots,N$ denotes the vertex linked to $a$ 
by the edge under consideration, with the convention that external 
edges corresponds to $b=0$;
\item  $j=1,\ldots,N_{ab}$ 
numbers the different edges between $a$ and $b$, 
$N_{ab}$ being their total number. We set $N_{ab}=0$ if 
$a$ is not connected to $b$.
\end{itemize}
In this way the ordered 
triplet $(a,b,j)$ uniquely defines all the oriented edges of the graph. 
Obviously, $(a,b,j)$ and $(b,a,j)$ define the same 
edge, but with a different orientation. Hence we have 
$N_{ab}=N_{ba}$. We will call internal mode (resp. 
external mode) a mode living on an internal edge (resp. external edge). The 
length of the edge $(a,b,j)$ is denoted $d_j^{ab}=d_j^{b a}$.
We denote $\cA_j^{ab}(x)$ the projection of the ambient vector potential 
$\mathbf{A}(\mathbf{x})$ on the unit tangent vector 
$\mathbf{t}_j^{ab}(x)$
of the edge $(a,b,j)$ chosen in accordance with the orientation of the edge and with $x\in[0,d_j^{ab}]$ being the 
variable of the parametrization of the embedded edge. As explained
in \cite{KS1}, the physical motivation of seeing the graph embedded in $\RR^3$ with an ambient vector potential is convenient in 
introducing the $\cA_j^{ab}$'s but in the sequel we may as well consider that these functions are given to us as part of the 
graph data. The effect of the external potential is now encoded easily in our formalism. One simply has to modify the relations 
describing the propagation of a mode $\fa_{j}^{ab}(k)$ (eq (2.2) in \cite{CR}) to
\bea
\label{propagation_magnetic}
\fa_{j}^{ab}(k) = 
\exp(-ik\,d^{ab}_{j}+i\theta_j^{ab})\,\fa_{j}^{b a}(-k)\,,
\eea
where 
\bea
\theta_j^{ab}=\int_0^{d_j^{ab}}\cA_j^{ab}(x)\,dx\,.
\eea
To see this, we simply note that instead of using the standard eigenfunctions of the operator $\frac{d}{dx}$ \ie $e^{-ikx}$, we now use 
the following eigenfunctions of the operator $\frac{d}{dx}-i\cA_j^{ab}(x)$ on the edge $(a,b,j)$: $e^{-ikx+
i\int_0^x\cA_j^{ab}(t)\,dt}$.
The consistency of (\ref{propagation_magnetic}) is ensured by the fact that $d_j^{ab}=d_j^{ba}$ and the property
\bea
\theta_j^{ab}=-\theta_j^{ba}\,,
\eea
which in turn follows from the fact that $\mathbf{t}_j^{ab}(x)=-\mathbf{t}_j^{ba}(x)$.

In the presence of a magnetic field, it is important to discuss loops and cycles. In this paper, we call a loop an internal edge that
connects a vertex to itself. A cycle is a closed simple path in the graph \ie a sequence of vertices such that each two 
adjacent vertices in the sequence are connected by an edge, and each two nonadjacent vertices in the sequence are not connected by any 
edge, with the starting and ending edge in the sequence being the same. These are sometimes also called simple cycles or chordless cycles or
polygons or holes. Note that a cycle of length $1$ is a loop. For a cycle of length $n>1$, denoting $\theta_1,\dots,\theta_n$ the 
line integrals of the $\cA$'s corresponding to the edges involved in the cycle and with the appropriate orientation, we see that 
the sum of the $\theta_j$'s is the flux of the magnetic field through the cycle by Stokes theorem. For a loop, we adapt the 
formalism of Section 3 in \cite{CR} along the same lines as above by defining\footnote{For convenience and conciseness, we add a variable $\theta$ 
in the arguments of $e_j^{ab}$
to account for the presence of the line integrals $\theta_j^{ab}$.}
\bea
e_j^{ab}(k,\theta)=\begin{cases}
e^{-ikd_j^{ab}+i\theta_j^{ab}}~~,~~\text{if}\, b\neq a\,\text{and }\,
 N_{ab}\neq 0\,,\\
e^{-ikd_j^{aa}}\left(\begin{array}{cc}
0&e^{i\theta_j^{aa}}\\
e^{-i\theta_j^{aa}}&0
\end{array}\right)~~,~~\text{if}\,b=a\,\text{and }\, 
N_{aa}\neq 0\,,
\end{cases}
\eea
where $\theta_j^{aa}=\int_0^{d_j^{aa}}\cA_j^{aa}(t)\,dt$ is the flux of the magnetic field through the loop, and 
then by writing the propagation relations as
\be\label{eq:prop-rel}
\fa_{j}^{ab}(k) = e_j^{ab}(k,\theta)\,\fa_{j}^{ba}(-k)\qquad
\forall j=1,\ldots,N_{ab}\ ;\ \forall b=0,1,\ldots,N\,.
\ee

\subsubsection*{Matrix form}
As in \cite{CR}, one can present the above relations in matrix form. However, since the 
determination of the sizes of the involved matrices is rather delicate, we prefer to recall here the main steps of \cite{CR}. 

We first look at the modes at a given vertex $a=1,...N$. One wishes to gather the internal modes $\fa_{j}^{ab}(k)$ ($b\neq 0$) 
in a vector $B_a(k)$ of \textit{minimal} size. For a given vertex $a$, let 
$M_a$ be the number of vertices connected to $a$, and $\{a_1, a_2, ... , a_{M_a}\}$ be these vertices (\ie $N_{a,a_i}\neq 0$ and $N_{ab}=0$ for $b\neq a_i$
, $i=1,\dots, M_a$). Let us denote by $v$ a generic basis vector of $\CC^p$ with $p$ being the appropriate dimension according to the context (indicated 
in the formulas below). The modes are then arranged into the vector $B_a$ as
\beq
B_a(k)=\sum_{\alpha=1}^{M_a}\sum_{j=1}^{N_{a,a_\alpha}} 
\underbrace{v_{\alpha}}_{M_a}\otimes \underbrace{v_j}_{N_{aa_\alpha}}\ \fa_{j}^{a,a_\alpha}(k)
\equiv \sum_{b=1}^{N}\sum_{j=1}^{N_{ab}} 
\underbrace{v_{b}}_{M_a}\otimes \underbrace{v_j}_{N_{ab}}\ \fa_{j}^{ab}(k)
\eeq
The first expression of $B_a(k)$ shows clearly the minimal structure (where we indicated the size of the basis vectors $v$)
while the second expression, easier to use in the following, has to be handled with the understanding that there is no slot created in $B_a$ when 
$N_{ab}=0$. The indicated sizes in this second expression are there to remind this. 

The full set of modes is then gathered in the vector
\beq
\cB(k)=\sum_{a=1}^N \underbrace{v_a}_{N}\otimes\ B_a(k)\,,
\eeq
which has $M_1+\dots+M_N$ components.

Then, the influence of the  magnetic field is entirely encoded in the following matrix 
\bea
\label{def_E}
E(k,\theta) = \sum_{a,b=1}^N 
\sum_{j=1}^{N_{ab}} \underbrace{E_{ab}}_{N\times N}\otimes 
\underbrace{E_{ba}}_{M_a\times M_a}\otimes \underbrace{E_{jj}}_{N_{ab}\times N_{ab}}\otimes\ 
e_j^{ab}(k,\theta)
\eea
where the matrix $E_{mn}$ is the square matrix with $1$ at position $(m,n)$ and $0$ elsewhere and appropriate dimension (given by the context and indicated 
explicitely in (\ref{def_E})). 
 In the following sections, we will consider "simple" graphs with $N_{ab}=1$ ($a\neq b$) and $N_{aa}=0$, so that the matrices 
$E_{jj}$ above will just drop out. An example (corresponding to a ring with $N$ external edges) of matrix $E(k,\theta)$ is given in the proof of 
proposition \ref{prop:eigenvalue}.

The relations \eqref{eq:prop-rel} are then totally encoded in the following relation
\beq
\label{eq:B}
\cB(k) = E(k,\theta)\,\cB(-k).
\eeq

The matrix $E(k,\theta)$ is the crucial ingredient in the total scattering matrix and enjoys the following important properties
\bea
E(k,\theta)^\dagger&=&E(-k,\theta)=E(k,\theta)^{-1} ,\quad \theta,\,k\in\RR\,,\\
\label{breaking}
E(k,\theta)^T&=&E(k,-\theta)\,,
\eea
where the notation $E(k,-\theta)$ means that all the $\theta_j^{ab}$ are changed to $-\theta_j^{ab}$ in $E$.
 One collects in a similar manner all the external modes $\fa_{j}^{a0}(k)$ in a vector $\mathfrak{A}(k)$ of size 
$N_{10}+\dots+N_{N0}$. The relations of local scattering at each vertex  read
\be
\label{local_comp}
\fa_{j}^{ab}(k) = 
\sum_{c=0}^N\sum_{\ell=1}^{N_{ac}} 
\fs^{bc}_{a;j\ell}(k)\, \fa_{\ell}^{ac}(-k)
\qquad\forall j=1,\ldots,N_{ab}\ ;\ \forall b=0,1,\ldots,N
\ee
where $\fs^{\beta\gamma}_{\alpha;jk}(p)$ are the components of the local 
scattering matrix 
$S_a(k)$. These relations are then gathered into
\bea
\label{eq:A1}
\mathfrak{A}(k) &=&
S^{\otot}(k)\,\mathfrak{A}(-k)+ S^{\otin}(k)\,\cB(-k)
\\
\label{eq:A2}
\cB(k) &=&
S^{\inot}(k)\,\mathfrak{A}(-k)+ S^{\inin}(k)\,\cB(-k)
\eea
where 
\bea
\label{block1}
S^{\otot}(k)&=& \sum_{a=1}^N
\sum_{j,\ell=1}^{N_{a0}} \underbrace{E_{aa}}_{N\times N}\otimes \underbrace{E_{j\ell}}_{N_{a0}\times N_{a0}}\otimes 
\fs^{00}_{a;j\ell}(k)\,,
\\
\label{block2}
S^{\otin}(k) &=& \sum_{a,c=1}^N
\sum_{j=1}^{N_{a0}}\sum_{\ell=1}^{N_{ac}}
\underbrace{E_{aa}}_{N\times N}\otimes
\underbrace{v_{c}^T}_{M_a}\otimes \underbrace{E_{j\ell}}_{N_{a0}\times N_{ac}}\otimes \fs^{0c}_{a;j\ell}(k)\,,
\\
\label{block3}
S^{\inot}(k) &=&\sum_{a,b=1}^N 
\sum_{j=1}^{N_{ab}}\sum_{\ell=1}^{N_{a0}}
\underbrace{E_{aa}}_{N\times N}\otimes
\underbrace{v_{b}}_{M_a}\otimes \underbrace{E_{j\ell}}_{N_{ab}\times N_{a0}}\otimes \fs^{b0}_{a;j\ell}(k)\,,
\\
\label{block4}
S^{\inin}(k) &=& \sum_{a,b,c=1}^N 
\sum_{j=1}^{N_{ab}}\sum_{\ell=1}^{N_{ac}}
\underbrace{E_{aa}}_{N\times N}\otimes
\underbrace{E_{bc}}_{N_a\times N_a}\otimes \underbrace{E_{j\ell}}_{N_{ab}\times N_{ac}}
\otimes \fs^{bc}_{a;j\ell}(k)\,.\quad
\eea
The explicit form of these matrices when considering the particular case of a ring with $N$ external edges is given in the proof of proposition \ref{prop:eigenvalue}.

Eliminating the internal modes $\cB(k)$ from equations (\ref{eq:B}), (\ref{eq:A1}) and (\ref{eq:A2}), 
 one obtains the expression of the total scattering matrix of the given graph 
\be
\label{expression_Stot}
\bS(k,\theta)=S^{\otot}(k)+S^{\otin}(k)\,\left[E(k,\theta)-S^{\inin}(k)\right]^{-1}
\,S^{\inot}(k)\,.
\ee
This formula applies to a completely general finite, connected, noncompact graph (\ie even with tadpoles and multiples edges between two vertices). 
For practical calculations in scattering problems, it is more amenable than the general results obtained in \cite{KS0}. Note that a similar formula 
appeared in the context of chaotic on graphs in \cite{KoSm} in the special case of a single edge between any two vertices and of symmetric local scattering 
matrices.

We can now see in detail an important general fact: 
the presence of a magnetic field breaks time-reversal invariance. It is known that time-reversal
invariance is equivalent to having a symmetric scattering matrix. So the fundamental source of time-reversal invariance breaking is 
equation (\ref{breaking}). Indeed, even when all the local scattering matrices are symmetric, hence giving 
${S^{\otin}}(k)^T={S^{\inot}}(k)$, ${S^{\inin}}(k)^T={S^{\inin}}(k)$ and ${S^{\otot}}(k)^T={S^{\otot}}(k)$, we still get
\be
\bS(k,\theta)^T=\bS(k,-\theta)\,.
\ee
The total scattering matrix is no longer symmetric in the presence of a magnetic field.

\subsection{Example of a quantum ring with $N$ external edges}\label{quantum_ring}

\subsubsection{Explicit form of the scattering matrix}

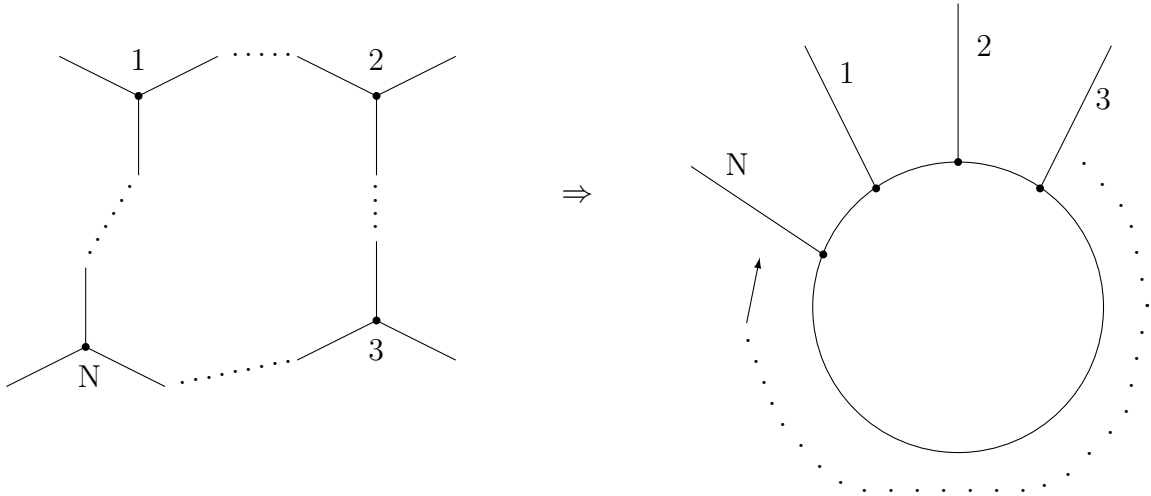
\begin{figure}[ht]
\begin{center}
\begin{picture}(400,200)
\put(20,150){\vertex{1}} \multiput(65,175)(5,0){5}{.}\put(110,150){\vertex{2}}
\multiput(118,125)(0,-5){4}{.}
\put(110,60){\vertexD{3}} \put(0,50){\vertexD{N}}
\multiput(85,59)(-5,-1){9}{.}
\multiput(10,100)(3,5){6}{.}
\put(190,120){$\Rightarrow$}
 \put(340,80){\circle{110}} 
\put(289,100){\line(-3,2){50}} \put(289,100){\circle*{3}} \put(252,130){N} 
\put(309,125){\line(-1,2){27}} \put(309,125){\circle*{3}}  \put(295,165){1}    
\put(340,135){\line(0,1){60}} \put(340,135){\circle*{3}} \put(347,175){2}
\put(371,125){\line(1,2){27}} \put(371,125){\circle*{3}}  \put(392,155){3}  
\multiput(410,80)(-2,10){4}{.}
\multiput(400,118)(-7,8){3}{.}
\multiput(410,80)(-2,-10){5}{.}
\multiput(396,32)(-8,-8){3}{.}
\put(371,12){.}
\multiput(363,10)(-10,0){7}{.}
\put(294,13){.}
\multiput(286,19)(-7,8){3}{.}\multiput(268,43)(-3,8){4}{.}
\put(260,74){\vector(2,10){5}}
%
\end{picture}
\caption{Gluing of $N$ three-edge identical star-graphs to obtain a regular ring.} 
\end{center}
\label{gluing}
\end{figure}

We apply the general method for computing the total scattering to a regular ring with $N$ external edges obtained by 
gluing $N$ three-edge vertices (see Fig. \ref{gluing}). 
            
As explained in \cite{CR}, the total scattering matrix is always 
defined up to a permutation accounting for the freedom in choosing a numbering and labelling of the vertices and edges.
For convenience we number the vertices clockwise
and we arrange the internal modes clockwise too, starting from the the external mode $\fa^{a 0}(k)$, $a=1,...,N$ at each vertex.
The other two modes at a vertex are\footnote{Of course, at vertex 1, the other modes are $\fa^{12}(k)$ and $\fa^{1N}(k)$, while at vertex $N$ they are $\fa^{N1}(k)$ and $\fa^{N,N-1}(k)$.} $\fa^{a,a+1}(k)$ and $\fa^{a,a-1}(k)$. 

Note that we drop the index $j$ for convenience as
there is only one edge between any two adjacent vertices. 

We assume that the internal edges are all of the same length 
$d$ and identify $N+1$ with $1$ for later convenience. The local scattering matrices are all taken to be the 
same $3\times 3$ matrix $\sloc(k)$, 
 satisfying the usual properties 
\bea
\sloc(k)\sloc(-k)=\1~~,~~\sloc(k)\sloc^\dagger(k^*)=\1\,.
\eea

To help the reader with notations and make the connection with the previous section, at vertex  $a=1,\dots,N$ we write
\bea
\left(\begin{array}{c}
\fa^{a 0}(k)\\
\fa^{aa+1}(k)\\
\fa^{aa-1}(k)
\end{array}\right)
&=& \sloc(k)\left(\begin{array}{c}
\fa^{a 0}(-k)\\
\fa^{aa+1}(-k)\\
\fa^{aa-1}(-k)
\end{array}\right) =
\left(\begin{array}{ccc} S_{11}(k) & S_{12}(k) & S_{13}(k) \\
S_{21}(k) & S_{22}(k) & S_{23}(k) \\
S_{31}(k) & S_{32}(k) & S_{33}(k) 
\end{array}\right)\left(\begin{array}{c}
\fa^{a 0}(-k)\\
\fa^{aa+1}(-k)\\
\fa^{aa-1}(-k)
\end{array}\right)
\nonu &=&
\left(\begin{array}{ccc} \fs^{00}_{a,11}(k) & \fs^{0,a+1}_{a,11}(k) & \fs^{0,a-1}_{a,11}(k) \\
\fs^{a+1,0}_{a,11}(k) & \fs^{a+1,a+1}_{a,11}(k) & \fs^{a+1,a-1}_{a,11}(k) \\
\fs^{a-1,0}_{a,11}(k) & \fs^{a-1,a+1}_{a,11}(k) & \fs^{a-1,a-1}_{a,11}(k) 
\end{array}\right)\,\left(\begin{array}{c}
\fa^{a 0}(-k)\\
\fa^{aa+1}(-k)\\
\fa^{aa-1}(-k)
\end{array}\right)\,.
\label{eq:Sloc}
\eea
The second line of \eqref{eq:Sloc} uses the notation of the general framework depicted in section \ref{sect:general}, while
the first line of \eqref{eq:Sloc} uses a simplified notation, allowed by the particular case we consider in this section.

\begin{figure}[ht]
\begin{center}
\begin{picture}(400,200)
 \put(90,80){\circle{110}} \put(88,78){$\Phi$} 
\put(39,100){\line(-3,2){50}} \put(39,100){\circle*{3}}  \put(2,130){N} \put(10,90){{\small{$S(k)$}}}
\put(59,125){\line(-1,2){27}} \put(59,125){\circle*{3}}  \put(45,165){1} \put(30,125){{\small{$S(k)$}}}
\put(90,135){\line(0,1){60}} \put(90,135){\circle*{3}} \put(97,175){2} \put(63,140){{\small{$S(k)$}}}
\put(121,125){\line(1,2){27}} \put(121,125){\circle*{3}}  \put(142,155){3} \put(100,137){{\small{$S(k)$}}}
\multiput(170,80)(-2,10){4}{.}
\multiput(160,118)(-7,8){3}{.}
\multiput(170,80)(-2,-10){5}{.}
\multiput(156,32)(-8,-8){3}{.}
\put(131,12){.}
\multiput(123,10)(-10,0){7}{.}
\put(54,13){.}
\multiput(46,19)(-7,8){3}{.}\multiput(28,43)(-3,8){4}{.}
\put(20,72){\vector(1,10){1}}
\put(200,80){$\Rightarrow$}
 \put(320,80){\circle*{10}} \put(290,60){$\bS(k,\theta)$} 
 \put(320,80){\line(-3,2){50}} \put(260,100){N}
 \put(320,80){\line(0,1){60}} \put(310,130){1}
 \put(320,80){\line(3,-2){50}} \put(350,110){2}
 \put(320,80){\line(3,2){50}} \put(350,65){3}
\multiput(350,34)(-8,-8){3}{.}
\put(327,14){.}
\multiput(320,12)(-10,0){3}{.}
\put(290,15){.}
\multiput(282,21)(-7,8){3}{.}\multiput(264,45)(-3,8){4}{.}
\put(256,74){\vector(1,10){1}}
\end{picture}
\end{center}
\caption{Total scattering matrix of the regular ring.} 
\label{total}
\end{figure}
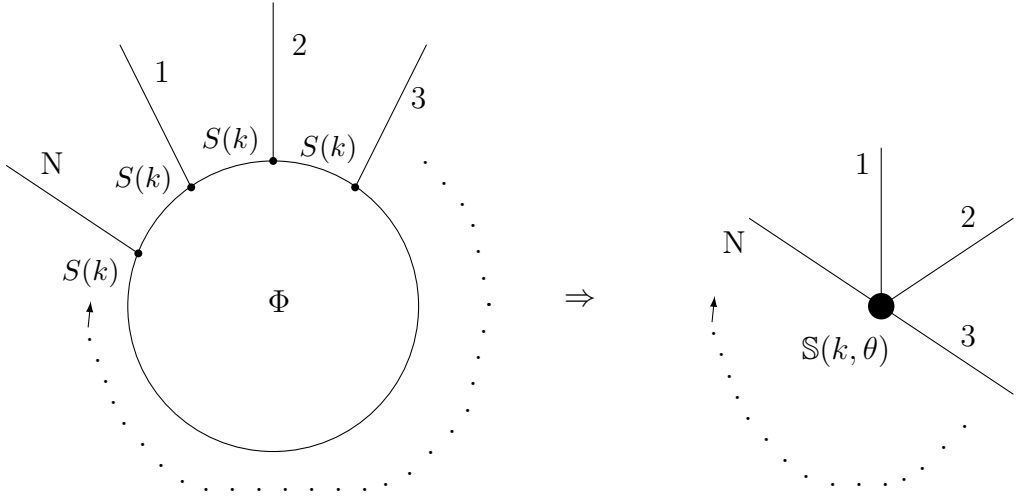

We assume that the magnetic field $B$ is 
constant and uniform along the direction normal to the plane containing the ring. Setting up a right-handed set of axis $(Ox,Oy,Oz)$ with some fixed 
origin $O$ and such that the polygon lies in the $(x,y)$ plane, $B$ is taken in the positive $z$ direction. 
Given the symmetry of the problem, 
the line integrals of the corresponding vector potential along the edge relating $a$ to $a+1$ are all equal to 
$\frac{\Phi}{N}$ where $\Phi$ is the (negative) total flux of $B$ through 
the surface of the ring. We denote $\theta=-\frac{\Phi}{N}$ the "flux per edge".
It is then clear that there is a cyclic structure on this graph (see Fig. \ref{total}). 

\begin{proposition}\label{prop:eigenvalue}
The total scattering matrix $\bS(k,\theta)$ of the quantum ring with $N$ external edges and pierced by the flux $\Phi$
 is a circulant matrix and can be diagonalized as 
\bea
\bS(k,\theta)=W^{-1}\Lambda(k,\theta)W\,,
\eea
where 
\bea
W=\frac{1}{\sqrt{N}}\sum_{a,b=1}^N \omega^{(a-1)(b-1)}E_{ab}=\frac{1}{\sqrt{N}}\left(\begin{array}{cccc}
1&1&\dots &1\\
1&\omega&\dots & \omega^{N-1}\\
\vdots& \vdots&\ddots& \vdots\\
1&\omega^{N-1}& \dots& \omega
\end{array}\right)~~,~~\omega=e^{\frac{2i\pi}{N}}\,,
\eea
and 
\bea
\Lambda_{ab}(k,\theta)=\delta_{ab}\,\lambda\left(k,\theta+\frac{2(b-1)\pi}{N}\right)~~,~~a,b=1,\dots,N\,.
\eea
The eigenvalue $\lambda(k,\theta)$ is called the fundamental eigenvalue and reads
\bea
\lambda(k,\theta)=\frac{e^{2ikd}\det \sloc+e^{ikd}\left[(\sloc_{11}\sloc_{23}-\sloc_{13}\sloc_{21})
e^{i\theta}+
(\sloc_{11}\sloc_{32}-\sloc_{31}\sloc_{12})e^{-i\theta}\right]-\sloc_{11}}
{e^{2ikd}(\sloc_{22}\sloc_{33}-\sloc_{23}\sloc_{32})+e^{ikd}(\sloc_{23}e^{i\theta}+\sloc_{32}e^{-i\theta})-1}\,,
\label{fundEig}
\eea
where we have dropped the $k$-dependence in the entries $S_{ab}$ for conciseness.

Let us stress that the above formulas are valid whatever the local matrix $S(k)$ 
(provided it is the same at each vertex).
\end{proposition}

\prf
Denote by $E^{P}_{ab}$ the matrix 
of size $P\times P$ with $1$ at position $(a,b)$. Then, the $2N\times 2N$ matrix $E(k,\theta)$ reads
\bea
E(k,\theta)&=&e^{-ikd}\sum_{a=1}^N\left(e^{-i \theta}E^{N}_{a,a+1}\otimes E^2_{12}+e^{i \theta}E^N_{a+1,a}
\otimes E^2_{21}\right)\,\\
&=&e^{-ikd}\sum_{a=1}^{N}\left(e^{-i \theta}E^{2N}_{2a-1,2a+2}+e^{i \theta}E^{2N}_{2a+2,2a-1}
\right)\,.
\eea
Denoting $\sloc_{ab}$, $a,b=1,2,3$ the elements of the local scattering matrix $\sloc(k)$, 
we obtain the four matrices $S^{\otot}(k)$, $S^{\otin}(k)$, $S^{\inin}(k)$, $S^{\inot}(k)$ that correspond to connection between external 
(out) or internal (in) edges. Explicitly, they read
\bea
S^{\otot}(k)=
\left(\begin{array}{ccc}\sloc_{11}&&\\&\ddots&\\&&\sloc_{11}\end{array}\right)=\sloc_{11}\1_N\,,
\eea
\bea
S^{\inin}(k)
&=&\left(\begin{array}{cccccccc}
\sloc_{22}&\sloc_{23}& & & & & & \\
\sloc_{32}&\sloc_{33}& & & & & & \\
& & \sloc_{22}&\sloc_{23} & & & & \\
& & \sloc_{32}&\sloc_{33} & & & & \\
& & & &\ddots & & & \\
& & & & &\ddots & & \\
& & & & & &\sloc_{22} &\sloc_{23}\\
& & & & & &\sloc_{32} &\sloc_{33}\\
\end{array}\right)\,,\\
&=&\1_N\otimes\left(\begin{array}{cc}
\sloc_{22}&\sloc_{23}\\
\sloc_{32}&\sloc_{33}\\
\end{array}\right)\,,
\eea
and
\bea
S^{\inot}(k)
&=&\left(\begin{array}{cccccccc}
\sloc_{21}& & & & & & & \\
\sloc_{31}& & & & & & & \\
& & \sloc_{21}& & & & & \\
& & \sloc_{31}& & & & & \\
& & & &\ddots & & & \\
& & & & & & &\sloc_{21}  \\
& & & & & & &\sloc_{31}
\end{array}\right)
\ =\ \1_N\otimes\left(\begin{array}{c}
\sloc_{21}\\ \sloc_{31}
\end{array}\right) \\[1.2ex]
&=&{S^{\otin}}(k)^T\,.
\eea
\bea
S^{\otin}(k)
&=&\left(\begin{array}{cccccccc}
\sloc_{12}&\sloc_{13}& & & & & & \\
& & \sloc_{12}&\sloc_{13} & & & & \\
& & & &\ddots & & & \\
& & & & & &\sloc_{12} &\sloc_{13}\\
\end{array}\right)\,,\\
&=&\1_N\otimes\left(\begin{array}{cc}
\sloc_{12}&\sloc_{13}\\
\end{array}\right)\,.
\eea

One can see that $\bS(k,\theta)$ is a circulant matrix by noticing that it commutes with the $N\times N$ shift operator
\bea
D=E^{N}_{N1}+\sum_{a=1}^{N-1} E^{N}_{a,a+1}~~\text{with}~~D^N=\1_N\,.
\eea
This follows immediately from 
\bea
&& DS^{\otot}=S^{\otot}D\,,\ DS^{\otin}=S^{\otin}\Delta\,,\ S^{\inot}D^{-1}=\Delta^{-1}S^{\inot}
\qquad
\\
&&\mb{and}
\Delta\big(E(k,\theta)-S^{\inin}(k,\theta)\big)=\big(E(k,\theta)-S^{\inin}(k,\theta)\big)\Delta\,,
\eea
 where $\Delta=D\otimes \1_2$. Therefore, $\bS(k,\theta)$ is 
diagonalizable by the matrix $W$. We write 
\bea
\label{S_diago}
\bS(k,\theta)=W^{-1}\sum_{a=1}^N\Big(\lambda_a(k,\theta)\,E^N_{aa}\Big) W\,.
\eea
Next, noting that 
\bea
WS^{\otin}=S^{\otin}(W\otimes \1_2)~~,~~S^{\inot}W^{-1}=(W^{-1}\otimes \1_2)S^{\inot}\,,
\eea
and
\be
(W\otimes \1_2)\big(E(k,\theta)-S^{\inin}\big)(W\otimes \1_2)^{-1}=\sum_{a=1}^NE^N_{aa}\otimes 
\left(\begin{array}{cc}
-\sloc_{22} & e^{-ikd-i\theta}\omega^{1-a}-\sloc_{23}\\
e^{-ikd+i\theta}\omega^{a-1}-\sloc_{32} & -\sloc_{33}
\end{array}
\right)
\ee
and inserting in (\ref{expression_Stot}), we obtain
\be
\lambda_a(k,\theta)=\frac{e^{2ikd}\det \sloc+e^{ikd}\left[(\sloc_{11}\sloc_{23}-\sloc_{13}\sloc_{21})e^{i\theta}\omega^{a-1}+
(\sloc_{11}\sloc_{32}-\sloc_{31}\sloc_{12})e^{-i\theta}\omega^{1-a}  \right]-\sloc_{11}}
{e^{2ikd}(\sloc_{22}\sloc_{33}-\sloc_{23}\sloc_{32})+e^{ikd}(\sloc_{23}e^{i\theta}\omega^{a-1}+\sloc_{32}e^{-i\theta}\omega^{1-a})-1}\,.
\ee
The final result follows from the observation
\be
\lambda_a(k,\theta)=\lambda_1\big(k,\theta+\frac{2(a-1)\pi}{N}\big)\,,
\ee
and setting $\lambda_1=\lambda$.
\finprf

\subsubsection{Miscellaneous properties of $\bS(k,\theta)$:}

Note that the property $\sloc(k)\sloc(-k)=\1_3$ allows us to give a more symmetric form of the 
fundamental eigenvalue
\bea
\lambda(k,\theta)=-\det \sloc(k)\ \frac{g(k,\theta)}{g(-k,\theta)}\,,
\eea
where 
\bea
g(k,\theta)=e^{ikd}-\big(\sloc_{23}(-k)e^{i\theta}+\sloc_{32}(-k)e^{-i\theta}\big)
-\det \sloc(-k)\,\sloc_{11}(k)e^{-ikd}\,.
\eea
The unitarity of $\bS(k,\theta)$ is equivalent to the relations
\bea
g^*(k^*,\theta)=g(-k,\theta)\,.
\eea
In particular we have $\lambda^*(k^*,\theta)\lambda(k,\theta)=1$ as required.

Consequently, another interesting form is the following discrete Fourier type representation
\bea
\bS(k,\theta)=\frac{1}{N}\sum_{j,\ell=0}^{N-1} \lambda\big(k,\theta+\frac{2j\pi}{N}\big)e^{-i\frac{2\pi j\ell}{N}}D^{\ell}\,.
\eea
Finally, we note the quasi-periodicity of $\bS(k,\theta)$
\bea
\bS\big(k,\theta+\frac{2\pi}{N}\big)=\Omega^{-1}\, \bS(k,\theta)\,\Omega\,,
\eea
where $\Omega=diag(1,\omega,\dots,\omega^{N-1})$ with $\omega=e^{\frac{2 i\pi}{N}}$. In particular, the independent elements of $\bS(k,\theta)$ (e.g. the elements of the first row) satisfy
\bea
\left(\bS\big(k,\theta+\frac{2\pi}{N}\big)\right)_{1b}=\omega^{-(b-1)}\,\big(\bS(k,\theta)\big)_{1b}~~,~~b=1,\dots,N\,,
\eea
and it is sufficient to consider $\theta\in[0,\frac{2\pi}{N}]$.

In the important special case of a scale invariant local matrix, we see that the only dependence of 
$\bS(k,\theta)$ on $k$ is through the term $e^{ikd}$. Hence, in that case, $\bS(k,\theta)$ is periodic in its first argument with period $\frac{2\pi}{d}$
\bea
\forall\, z\in\CC~~,~~\bS(z,\theta)=\bS\big(z+\frac{2\pi}{d},\theta\big)\,.
\eea
Therefore, we only need to study the poles of $\bS(k,\theta)$ in the strip $-\frac{\pi}{d}\le Re(z)<\frac{\pi}{d}$.

\subsubsection{Properties of the fundamental eigenvalue}

In the second half of this paper, we want to apply our general framework to study the effect of the magnetic flux on transport properties of a
 quantum ring. One clear effect of the flux is that it discriminates between clockwise and anticlockwise rotation around the ring. 
 We want this effect to come solely from 
the flux and not from local details of the junctions of the ring to the external edges. So, from now on, we concentrate on local matrices 
satisfying
\bea
P\sloc(k) P^{-1}=\sloc(k)\,,
\eea
where 
\bea
P=\left(\begin{array}{ccc}
1&0 &0\\
0&0 &1\\
0 & 1&0
\end{array}\right)\,.
\eea
Of particular interest to us will be the following scale-invariant (independent of $k$) local scattering matrix
\bea
\label{local_Smat}
S^C= \left(\begin{array}{ccc}
1-2t & \sqrt{2t(1-t)} & \sqrt{2t(1-t)} \\
\sqrt{2t(1-t)} & t-1 & t \\
\sqrt{2t(1-t)} & t & t-1
\end{array}\right)\,,\mb{with} t\in[0\,,\,1]  \,.
\eea
The fundamental eigenvalue (\ref{fundEig}) then reads
\bea
\label{invariant_lambda}
\lambda(k,\theta)=-\frac{t(\cos\theta-\cos kd)+i(t-1)\sin kd}{t(\cos\theta-\cos kd)-i(t-1)\sin kd}\,.
\eea
The two special values $t=0$ and $t=1$ yield trivial results and will not be considered in the rest of this paper.
The following observation will have important consequences in the next section. If $\theta=0$ then 
\bea
\label{exp:lambda0}
\lambda(k,\theta)=1-i\frac{td}{t-1}\,k-\frac{d^2t^2}{2(t-1)^2}k^2+O(k^3)\equiv 1-i\chi_0\,k-\frac{\chi_0^2}{2}\,k^2+O(k^3) \,,
\eea
while, for $\theta\neq 0$, 
\beq
\label{exp:lambda-theta}
\lambda(k,\theta)=-1+i\frac{(t-1)d}{t\sin^2\frac{\theta}{2}}k+\frac{2(t-1)^2d^2}{t^2\sin^4\frac{\theta}{2}}k^2+O(k^3)\equiv 
-1+i\chi(\theta)\,k+2\chi(\theta)^2\,k^2+O(k^3)\,.
\eeq
In other words,
\bea
\lim_{k\to 0}\lim_{\theta\to 0}\lambda(k,\theta)\neq\lim_{\theta\to 0}\lim_{k\to 0}\lambda(k,\theta)\,.
\eea
The pole structure of 
$\bS(k,\theta)$ is controlled by the poles of its fundamental eigenvalue given by the equation
\bea
\label{BAE}
e^{-2ikd}-2\sloc_{23}\cos\theta e^{-ikd}
+\sloc_{23}^2-\sloc_{22}^2=0\,.
\eea
In general, it cannot be solved in closed form. When considering
$S^C$ they are given by
\bea
e^{-ikd} &=& t\cos\theta\pm\sqrt{\big(1-t(1-\sin\theta)\big)\big(1-t(1+\sin\theta)\big)}\\
&=& t\cos\theta\pm\sqrt{1-2t+t^2\cos^2\theta}
\eea
It admits real solution for $k$ only when $t=1$ or when $\cos\theta=\pm1$.

For $t\sim1$, it simplifies to $e^{-ikd}\sim e^{\pm i\theta}\mp(1-t)\sin\theta$.  When $\cos\theta=\pm1$, we have $kd=(2n+1)\frac\pi2$, $n\in\ZZ$.

\section{Physical models \label{phys_mod}}

\begin{figure}[ht]
\begin{center}
\begin{picture}(100,200)(160,-260)
\includegraphics[scale=0.55,angle=-90]{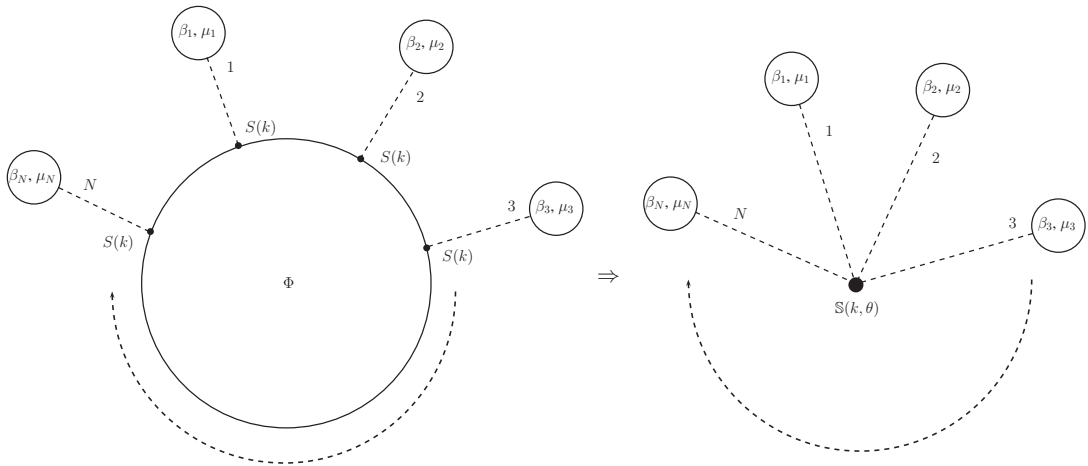}
\end{picture}
\caption{The ring and its effective representation as a star-graph connected to reservoirs.} 
\end{center}
\label{reservoirs}
\end{figure}

For applications to mesoscopic systems, we want to think of the previous formalism as a support to formulate theories of fermions (or bosons) 
propagating in (effectively) one-dimensional structures (the edges of the graph), being scattered at localized points in space (the vertices of the 
graph) and originating from reservoirs connected to the external edges of the graph (see Fig. \ref{reservoirs}). 
The modes $\fa_{j}^{ab}(k)$ 
of section \ref{general_method} are now seen as annihilation operators of particles with momentum $k$ and living on edge $(a,b,j)$. 
Corresponding creation operators $\fa_{j}^{ab\dagger}(k)$ are also considered. The point of view taken here is that one is only interested 
in the correlations and transport properties between the external edges of the graph. This is reminiscent of the B\"uttiker approach 
where the sample connected to the reservoirs is seen as a sort of black box entirely described by its scattering properties. 
The crucial new ingredient here is that we have detailed information on the internal structure of the black box: it is a 
graph whose metric structure and local scattering properties are known. This is what allows us to derive the total scattering matrix of our 
special sample. Finally, only the creation and annihilation operators corresponding to the states on the external edges are kept and 
they form a Reflection-Transmission algebra 
\cite{MRS} with the scattering matrix given by the total scattering matrix of the graph as calculated in the previous section. 

This formalism accommodates fermions and bosons and any dispersion relation of the underlying theory. In particular, nonrelativistic (Schr\"odinger) and 
relativistic (Dirac) systems are equally well and easily described. Note that special interactions amenable to bosonization techniques can also be 
studied this way. All these ideas have been studied in the context of the formulation of quantum field theory on a star-graph in a series of 
papers by one of the authors and collaborators \cite{Mintchevetal}. In the following, we argue that the main transport formulas derived in this way can be used directly for 
an arbitrary graph with magnetic field simply by using the total scattering matrix described in the previous section. In a sense, this represents a 
rigorous justification  of the traditional Landauer-B\"uttiker formalism using QFT methods. In particular, the incorporation of the magnetic field 
follows from a minimal coupling of the particle fields to an external vector potential.

 In the following, since we consider a ring connected to reservoirs, we will drop the index $j$, 
as in section \ref{quantum_ring}. Moreover, we will compute only quantities referring to external edges, the role of the internal edges being encoded in the total scattering matrix. Then, instead of 
labeling these external edges by $(a,0)$, as in the general formalism, we will simply use the index $a$.

\subsection{Schr\"odinger ring}

We recall the main transport formulas for our setup before applying them in the next section by taking full advantage of the fact 
that we have a complete analytic knowledge of the total scattering matrix of the quantum ring. For conciseness, we denote $\bS(k,\theta)$ by 
$\bS^\Phi(k)$ where $\theta=-\Phi/N$.

\subsubsection{Current} 

Using a direct adaptation of the explicit construction \cite{Mintchev:2011mx} of the steady state
 describing 
the quantum ring connected to the reservoirs with inverse temperature $\beta_a$ and chemical potential $\mu_a$, 
one finds the following Landauer-B\"uttiker \cite{la-57, bu-86} formula for the expectation value of the current operator
\begin{equation}
\langle j_x(t,x,a) \rangle_{\beta, \mu}  
=\int_0^\infty \frac{\rd k}{2\pi} \frac{k}{m} \sum_{b=1}^N \left [\delta_{ab} -  
|\bS^\Phi_{ab}(k)|^2\right ] d_b(k)\, , 
\label{LB}
\end{equation} 
where 
\begin{equation} 
d_a (k) = \frac{\e^{-\beta_a [\omega (k) - \mu_a]}}{1+ \e^{-\beta_a [\omega (k) - \mu_a]}} \, , 
\qquad \omega (k)=\frac{k^2}{2m},
\label{fbd1} 
\end{equation}
is the familiar Fermi distribution. Let us emphasize that the matrix elements $\bS^\Phi_{ab}(k)$ appearing in equation (\ref{LB}) are the ones for the 
total scattering matrix encoding the full information about the quantum ring, as computed in section \ref{quantum_ring}. 

The periodicity of $\bS^\Phi$ in $\Phi$, established in the previous section, indicates that the current oscillates 
with period $\Phi_0=2\pi$. The unitarity of $\bS^\Phi$ implies the following $t$-independent bound 
\begin{eqnarray}
\big|\langle j_x(t,x,a) \rangle_{\beta, \mu} \big| \leq 
\sum_{b\neq a}\left(
\frac{\log\left [1+ e^{\mu_a\beta_a}\right ]}{2 \pi \beta_a }
+\frac{\log\left [1+ e^{\mu_b\beta_b}\right ]}{2 \pi \beta_b }\right) \, . 
\label{est1}
\end{eqnarray}
From that expression, one sees that in the pure thermal regime $\mu_a=\mu\to 0$, the current vanishes (as it should).
However, as we show in the next section, its fluctuations, encoded in the thermal noise, do not. Moreover, the magnetic field has a non-trivial effect on the noise, as we now discuss.

\subsubsection{Noise}
The {\it zero frequency noise power} is defined as usual \cite{bb-00} by 
\begin{equation}
P_{ab} = \lim_{\nu \to 0^+} \int_{-\infty}^\infty d t\,  e^{i \nu t} \, 
\langle j_x(t,x_1,a) j_x(0,x_2,b) \rangle_{\beta, \mu}^{\rm conn}\, , 
\label{N1}
\end{equation}
where 
$\langle j_x(t_1,x_1,a) j_x(t_2,x_2,b) \rangle_{\beta, \mu}^{\rm conn}$ 
is the {\it connected} current-current correlation function in the state $\Omega_{\beta,\mu}$. 
It turns out \cite{Mintchev:2011mx} that $P_{ab}$ is $x_{1,2}$-independent and is given by 

\begin{eqnarray}
P_{ab} &=& \frac{1}{m} \int^{\infty}_0 \frac{d k}{2 \pi} k 
\Bigl \{ \delta_{ab} d_{a}(k) c_{a}(k) 
- |\bS^\Phi_{ab}(k)|^2 d_{b}(k) c_{b}(k) - |\bS^\Phi_{ba}(k)|^2 d_{a}(k) c_{a}(k) 
\nonumber \\
&&+\frac{1}{2}\sum_{e,f=1}^N \bS^\Phi_{ae}(k)\overline{\bS}_{be}^\Phi(k) 
\bS^\Phi_{bf}(k)\overline{\bS}^\Phi_{af}(k) [c_e(k)d_f(k)+c_f(k)d_e(k)]\Bigr \}\, , 
\qquad \qquad 
\label{N2}
\end{eqnarray}
where $c_a(k) \equiv 1-d_a(k)$.
$P$ is a symmetric matrix \cite{Mintchev:2011mx}. If we assume now $\mu_a=\mu$ and 
$\beta_a=\beta$, so that the setup of Fig. \ref{reservoirs} respects the cyclic symmetry of the ring, 
then $P$ is also a circulant matrix. Combining this fact 
with the Kirchhoff rule 
\begin{equation}
\sum_{b=1}^N P_{ab} = 0\, , \qquad a=1,2,...,N\, , 
\label{Kr2}
\end{equation}
we get $P_{ab}=P_{ba}=P_{|a-b|}$ with $P_n=P_{N-n}$, $n=0,1,..,N-1$, given by
\bea\label{eq:Pn}
P_{n}= \frac{2}{m} \int^{\infty}_0 \frac{d k}{2 \pi} k\,  \frac{e^{-\beta(\frac{k^2}{2m}-\mu)}}{\big(1+e^{-\beta(\frac{k^2}{2m}-\mu)}\big)^2} 
\Bigl \{ \delta_{n,0}- |\sigma_{n}(k,\theta)|^2 \Bigl \},
\eea
where 
\bea\label{eq:sigmaj}
&&N\,\sigma_n(k,\theta) \ =\ \sum_{\ell=1}^N \omega^{(\ell-1)n}\, \lambda(k,\theta+(\ell-1)\frac{2\pi}{N}).
\eea
Since $\lambda(k,\theta)$ is a phase (when $k$ and $\theta$ are real), we immediately get a bound
\beq
0\leq |\sigma_n(k,\theta)|^2\leq 1
\eeq
that implies 
\beq 
0\leq P_n\leq \frac{2}{1+e^{-\beta\mu}}\,\frac{1}{\pi\beta}\,.
\eeq

The general framework presented in this paper has been applied in \cite{CRM} to the case of a ring with $N=3$ edges. Numerics have shown a surprising small 
temperature behavior of the pure thermal noise. Here we provide an analytic treatment of this behaviour for general $N$, using the case $N=2$ as 
a detailed illustration.  We focus on the pure thermal noise 
both analytically. These 
features are specific of the scale-invariant case and we illustrate this fact by considering a model for energy-dependent local scattering on the ring.

\subsubsection{$N=2$ case}\label{N=2}

From the general treatment given above, we know the
 eigenvalues:
\bea
\lambda(k,\theta)&=& -\,\frac{t\big(\cos(kd) -\cos\theta \big)-i(1-t)\sin(kd)}{t\big(\cos(kd) -\cos\theta \big)+i(1-t)\sin(kd)}
\\
\lambda_2(k,\theta)&=& \frac{t\big(\cos(kd)+\cos\theta \big)-i(1-t)\sin(kd)}{t\big(\cos(kd)+\cos\theta \big)+i(1-t)\sin(kd)}
\eea
They lead to
\bea
&& \rho(k,\theta)\ =\ \frac{1}{4}\Big| \lambda(k,\theta)+\lambda_2(k,\theta)\Big|^2\ =\  
\frac{\big(t^2\cos^2(\theta)-(t-1)^2+(1-2t)\cos^2(kd)\big)^2}{n(k,\theta)\ n(k,\pi-\theta)}
\qquad\\
&& \tau(k,\theta)\ =\ \frac{1}{4}\Big| \lambda(k,\theta)-\lambda_2(k,\theta)\Big|^2 \ =\  
\frac{4t^2(t-1)^2\cos^2(\theta)\sin^2(kd)}{n(k,\theta)\ n(k,\pi-\theta)}\\
&& n(k,\theta) \ =\  t^2\big(\cos(\theta)+\cos(kd)\big)^2+(1-t)^2\sin^2(kd)
\\
&& \rho(k,\theta)+\tau(k,\theta) \ =\ 1
\eea
Note the invariance $\tau(k,-\theta)=\tau(k,\theta+\pi)=\tau(k,\theta)$, so that we focus on $\theta\in[0\,,\,\frac\pi2]$.

Again we get
\bea
&&\lim_{t\to1}\,\lim_{\cos(kd)\to\pm\cos(\theta)}\, \tau(k,\theta) = 1 \mb{and} 
\lim_{\cos(kd)\to\pm\cos(\theta)}\, \lim_{t\to1}\,\tau(k,\theta) = 0
\\
&&\lim_{t\to1}\,\lim_{\theta\to n\pi}\, \tau(k,\theta) = 1 \mb{and} 
\lim_{\theta\to n\pi}\, \lim_{t\to1}\,\tau(k,\theta) = 0
\eea

The  noise has the form (for $\mu_1=\mu_2\equiv\mu$ and 
$\beta_1=\beta_2=\beta$):
\bea
P=P_{11}(\theta,\beta,\mu)\,\left(\begin{array}{cc}  1& -1 \\ -1 & 1\end{array}\right)
\eea
with
\bea
P_{11}(\theta,\beta,\mu_1,\mu_2) &=&\frac1m
\int_0^\infty \frac{kdk}{2\pi}\,
2\,d(k,\beta,\mu)\,c(k,\beta,\mu)\, \tau(k,\theta)
\qquad
\eea
We analyse below the behaviour of the noise at small temperature (i.e. large $\beta$). 
Performing in $P_{11}$ a change of variable $\eps=\frac{z}{\beta}+\mu$ with $\eps=k^2$ (we set $m=\frac12$),  we get
\bea\label{eq:Pinz}
P_{11}(\theta,\beta,\mu) &=&\frac2{\beta}
\int_{-{\mu}{\beta}}^\infty \frac{dz}{2\pi}\,
\tau \left (\sqrt{\frac{z}{\beta}+\mu},\theta\right )\frac{e^z}{(1+e^z)^2}\,.
\eea
Depending on the vanishing of $\mu$, the behaviour will be different, and thus we divide our study in two parts: $\mu=0$ and $\mu$ finite. 

We start with $\mu=0$, and rewrite \eqref{eq:Pinz} as
\bea
\label{scale_Lambda}
P_{11}(\theta,\beta,0) &=&\frac2{\beta}
\int_{0}^{\Lambda} \frac{dz}{2\pi}\,
\tau\left (\sqrt{\frac{z}{\beta}},\theta\right )\frac{e^z}{(1+e^z)^2}+\frac2{\beta}\int_{\Lambda}^\infty \frac{dz}{2\pi}\,
\tau\left (\sqrt{\frac{z}{\beta}},\theta\right )\frac{e^z}{(1+e^z)^2}\nonu
&=& I_0+I_1,\qquad
\eea
where $\Lambda$ is a parameter. Since $0\leq\tau(x,\theta,\mu)\leq1$, we get an upper value for the second integral $I_1\leq \frac{1}{2\pi\beta}\frac{1}{1+e^\Lambda}$ that decays exponentially with $\Lambda$.
For large values of $\beta$ in $I_0$, $x=\frac{z}{\beta}$ is small, and we can expand $\tau(\sqrt x,\theta)$ around $x=0$. 
 We find
\bea
\tau\left (\sqrt{\frac{z}{\beta}},\theta\right ) &=&\frac{4(1-t)^2d^2}{t^2}\frac{cos^2(\theta)}{\sin^4(\theta)} \frac{z}{\beta} +... \mb{when} \theta\neq0
\\
\tau\left (\sqrt{\frac{z}{\beta}},0\right ) &=& 1-\frac{(1-2t)^2d^2}{4(1-t)^2t^2} \frac{z}{\beta} +...
\eea
Plugging this expansion into $I_0$, we get
\bea
I_0 &=& \frac{4(1-t)^2d^2}{t^2}\frac{cos^2(\theta)}{\sin^4(\theta)} \frac{1}{\pi\beta^2}\Big\{\ln(2)-\ln(1+e^{-\Lambda})-\frac{\Lambda}{1+e^{\Lambda}}\Big\}
+...\mb{for} \theta\neq0\quad\\
I_0 &=& \frac{1}{\pi\beta}\Big\{\half -\frac{1}{1+e^{\Lambda}}\Big\}+...
\mb{for} \theta=0
\eea
Sending $\Lambda\to\infty$, we find that 
\beq
P_{11}(\theta,\beta,0)\sim
\begin{cases} \frac{1}{\beta^2} \mb{when} \theta\neq0\,,\\
 \frac{1}{\beta} \mb{when} \theta=0\,.
 \end{cases}
\eeq
Now we consider the case $\mu\neq0$ and rewrite \eqref{eq:Pinz} as
\bea
P_{11}(\theta,\beta,\mu) &=&\frac2{\beta}
\int_{-\beta\mu}^{\beta\mu} \frac{dz}{2\pi}\,
\tau\left (\sqrt{\frac{z}{\beta}+\mu},\theta\right )\frac{e^z}{(1+e^z)^2}+\frac2{\beta}\int_{\beta\mu}^\infty \frac{dz}{2\pi}\,
\tau\left (\sqrt{\frac{z}{\beta}+\mu},\theta\right )\frac{e^z}{(1+e^z)^2}\nonu
&=&J_0+J_1,\qquad
\eea
Again we get an exponential decay for $J_1$:
\beq
J_1\leq \frac{1}{\pi\beta}\frac{1}{1+e^{\beta\mu}}
\eeq
while the expansion of $\tau$ in $J_0$ gives
\bea
J_0 &=& \frac{4(1-t)^2d^2}{t^2}\frac{cos^2(\theta)}{\sin^4(\theta)} \frac{\mu}{\pi\beta}\tanh\left (\frac{\beta\mu}{2}\right )+...
\mb{for} \theta\neq0
\label{eq:J0theta}\\
J_0 &=& \frac{1}{\pi\beta}\tanh\left (\frac{\beta\mu}{2}\right )+...
\mb{for} \theta=0\label{eq:J00}
\eea
Thus we get
\beq
P_{11}(\theta,\beta,\mu)\sim \frac{1}{\beta}\qquad \forall \theta
\eeq

The interpolation between these two regimes can be obtained by considering that $\mu$ scales as $\frac1\beta$, for instance with $\mu\beta=L$. 
Then, repeating the argument associated to (\ref{scale_Lambda}), one can derive again the $\frac1{\beta^2}$ behaviour for $\theta\neq 0$ and 
the $\frac1{\beta}$ behaviour for $\theta=0$, which is consistent with the limits of \eqref{eq:J0theta} and \eqref{eq:J00} in this regime.

\subsubsection{General $N$ case} \label{general_N}

The same analysis can be done and it goes as follows. First, let us count the number of independent entries in the noise matrix $P_{ab}$. Using 
the fact that it is a symmetric, circulant matrix such that each line (or column) adds up to zero (Kirchhoff rule), one finds that if $N=2p$ is even or 
$N=2p+1$ is odd then there are $p$ independent matrix elements, say $P_0,\dots, P_{p-1}$ in (\ref{eq:Pn}). In view of (\ref{eq:sigmaj}), 
the case $n=0$ must then be considered separately from the case $n\neq 0$ in the discussion of the small $k$ behaviour of $\sigma_n$ 
both for generic values of $\theta$ and for the specific values $\theta=-\frac{2n\pi}{N}$, $n\geq 0$ where one has to use the expansion 
\eqref{exp:lambda0} instead of \eqref{exp:lambda-theta}. Performing the calculations, one finds 
\bea
1-|\sigma_0(k,\theta)|^2&=&-\frac{C_0(\theta)^2}{N^2}k^2 +O(k^4) \mb{when} \theta\neq0
\\
1-|\sigma_0(k,0)|^2 &=& 4\frac{N-1}{N^2}-\frac{A_0^2}{N^2}\,k^2 +O(k^4)
\eea
and, for $n=1,\dots,p-1$,
\bea
|\sigma_n(k,\theta)|^2&=&\frac{C_n(\theta)^2}{N^2}k^2 +O(k^4) \mb{when} \theta\neq0
\\
|\sigma_n(k,0)|^2 &=& \frac{4}{N^2}+\frac{A_n^2}{N^2}\,k^2 +O(k^4)
\eea
where, for $n=0,\dots,p-1$,
\beq
A_n=\sum_{\ell=2}^N \omega^{n(\ell-1)}\chi\left ((\ell-1)\frac{2\pi}{N}\right )-\chi_0~~,~~C_n(\theta)=\sum_{\ell=1}^N \omega^{n(\ell-1)}\chi\left (\theta+(\ell-1)\frac{2\pi}{N}\right)\,,
\eeq
and $\chi_0$, $\chi(\theta)$ are defined in (\ref{exp:lambda0}) and (\ref{exp:lambda-theta}).
Therefore, all the conclusions obtained for $N=2$ for the case $\mu=0$ or $\mu\neq 0$ go over to the general $N$ case. For instance, 
for $\mu=0$, we find 
\bea
P_0(\theta,\beta,0)=\begin{cases} -\frac{C_0(\theta)^2\ln 2}{\pi N^2}\frac{1}{\beta^2}+O(\frac{1}{\beta^3}) \mb{when} \theta\neq0\,,\\
  \frac{2(N-1)}{\pi N^2}\frac{1}{\beta}+O(\frac{1}{\beta^2}) \mb{when} \theta=0\,.
 \end{cases}
\eea
And in general, we find that when $\mu=0$ or vanishingly small, $P_{n}$ behaves likes $1/\beta^2$ for 
$\theta\neq 0$ but recovers the usual behaviour in $1/\beta$ at $\theta=0$. When $\mu$ is a finite constant, $P_{n}$ behaves like $1/\beta$ 
for all values of $\theta$. Note that the case $N=3$ was studied numerically in details in \cite{CRM} and the present results confirm and 
complete the numerical study.

\subsection{Dirac ring}
In \cite{CRM}, some results analogous to the previous ones case were announced for the Dirac case. In this section, we perform the 
analytic study of the Dirac ring in the general $N$ case and present some numerical results for the case $N=3$ that mimic those presented in \cite{CRM} for the Schr\"odinger case.  
We find that the Dirac case shares the same small temperature peculiar behaviour as the Schr\"odinger case, albeit 
with a different power law which is a signature of the different underlying dispersion relation (relativistic as opposed to non-relativistic).

Let us consider now relativistic fermions. In this case,  
one finds \cite{Mintchev:2011mx}  the following steady current 
\begin{eqnarray} 
\langle j_x(t,x,a)\rangle_{\beta,\mu,\tmu} = 
\int_0^{\infty}\frac{d k}{2\pi} \sum_{b=1}^N\left [\delta_{ab} -  
|\bS^\Phi _{ab}(k)|^2\right ] [f_b(k)-\ft_b(k)] \, ,    
\label{fcurr1}
\end{eqnarray}
where 
\begin{equation}
f_a(k) = \frac{e^{-\beta_a(|k| -\mu_a)}}{1+e^{-\beta_a(|k| -\mu_a)}}\, , \qquad 
\ft_a(k) = \frac{e^{-\beta_a(|k| +\tmu_a)}}{1+e^{-\beta_a(|k| +\tmu_a)}} \, 
\label{ddpa}
\end{equation}
are  the Fermi distributions for particles and antiparticles.

The zero frequency noise power  is given by
\begin{eqnarray}
P_{ab} &=& \int^{\infty}_0 \frac{d k}{2 \pi}
\Bigl \{ \delta_{ab} F_{aa}(k)  
- |\bS^\Phi_{ab}(k)|^2 F_{bb}(k) - |\bS^\Phi_{ba}(k)|^2 F_{aa}(k) 
\nonu
&&\qquad+\frac{1}{2}\sum_{c,d=1}^N \bS^\Phi_{ac}(k)\overline{\bS}_{bc}^\Phi(k) 
\bS^\Phi_{bd}(k)\overline{\bS}^\Phi_{ad}(k) [F_{cd}(k)+F_{dc}(k)]\Bigr \}\, , 
\label{DN2}
\end{eqnarray}
where 
\begin{equation}
F_{ab}(k) =  f_{a}(k) [1-f_{b}(k)] + \ft_{a}(k) [1-\ft_{b}(k)] \, . 
\label{ff11}
\end{equation}  

\begin{figure}[ht]
\begin{center}
\includegraphics[scale=0.8]{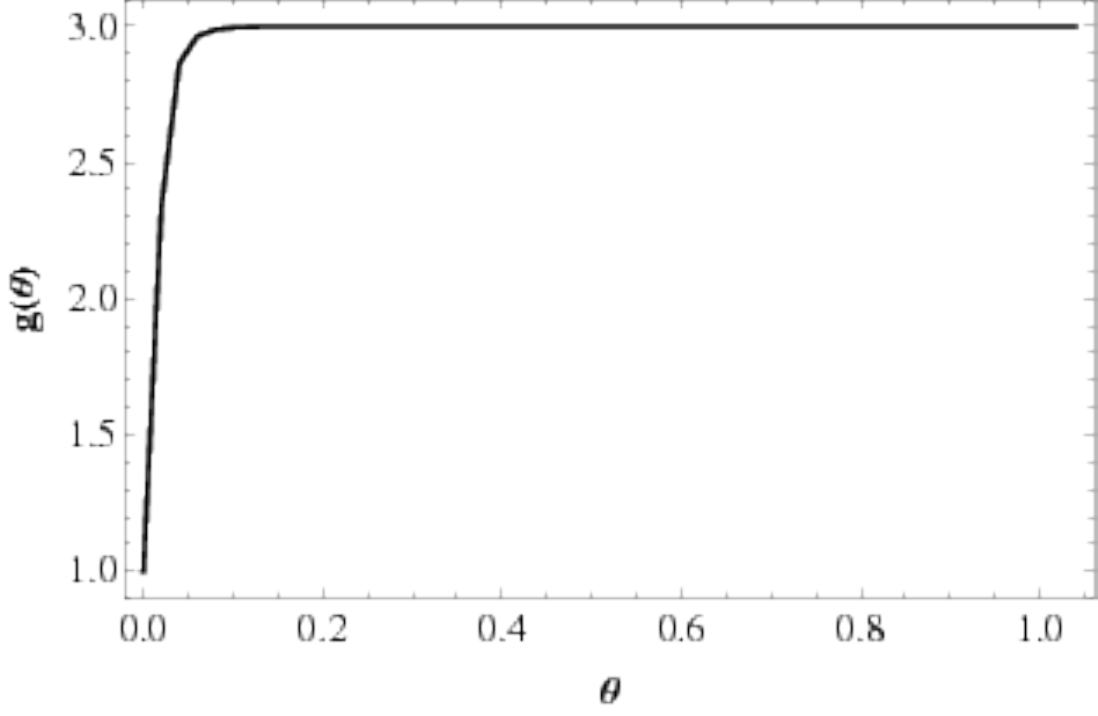}
\caption{Plot of the power $g(\theta)$ of $1/\beta$ as a function of $\theta\in[0,\frac{\pi}{6}]$ in the behaviour of $P^{Dirac}_{0}$.} 
\end{center}
\label{exponent_Dirac}
\end{figure}

Let us now discuss the case of the quantum ring with $N$ edges with the choice of $N$ identical local scattering matrices given 
by \cite{BMS}
\begin{equation} 
S(k) = \theta (k) U + \theta(-k) U^{-1} \, ,  
\label{DS1} 
\end{equation} 
 where $U$ is taken to be the matrix in (\ref{local_Smat}). In this case, the fundamental eigenvalue of the total scattering 
matrix is again given by (\ref{invariant_lambda}) and we can use all the relevant results of Section \ref{quantum_ring}.
Let us set $\mu_a=- \tmu_a= \mu$ and $\beta_a=\beta$ for $a=1,\dots,N$. The same argument as before leads to 
the following expression for the independent noise matrix elements $P^{Dirac}_n$, $n=0,\dots, p-1$ (where $N=2p$ or $N=2p+1$),
\bea
\label{eq:PnDirac}
P^{Dirac}_{n}= \frac{4}{\beta} \int^{\infty}_{-\beta\mu} \frac{d z}{2 \pi} \frac{e^{z}}{\big(1+e^{z}\big)^2} 
\Bigl \{ \delta_{n,0}- \Big|\sigma_{n}\Big(\frac{z}{\beta}+\mu,\theta\Big)\Big|^2 \Bigl \},
\eea
with $\sigma_n(k,\theta)$ given as before by (\ref{eq:sigmaj}). The main difference between the Dirac case and the Schr\"odinger case in our context 
is the dispersion relation which is linear in the Dirac case. This manifests itself in the previous equation by the fact that the first 
argument of $\sigma_n$ does not contain a square root. When we combine this fact with the arguments of Sections \ref{N=2} and \ref{general_N} which are 
still completely valid here, we obtain the following important results. When $\mu=0$ or vanishingly small, $P^{Dirac}_{n}$ behaves likes $1/\beta^3$ for 
$\theta\neq 0$ but recovers the usual behaviour in $1/\beta$ at $\theta=0$. When $\mu$ is a finite constant, $P^{Dirac}_{n}$ behaves like $1/\beta$ 
for all values of $\theta$.
The result of a numerical simulation performed for $N=3$ and $\mu=0$ and plotted in Fig. \ref{exponent_Dirac} confirms this statement. It also shows 
that there is an interpolation region between the two regimes. This interpolation region has been studied numerically in more details in \cite{CRM} 
(for the Schr\"odinger case) and shows that the "jump" from the value at $\theta=0$ to the plateau value ($3$ in the Dirac case) is smooth and follows 
a universal profile. An analytical treatment of this region remains an interesting open problem.
We see that, as in the Schr\"odinger case, magnetic fluxes $\Phi \not= n\Phi_0$ modify for $T\sim 0$ 
the usual linear temperature dependence of the pure thermal noise. In spite of the fact that 
the scattering matrix $\bS^\Phi$ for both Schr\"odinger and Dirac junctions is the same, 
the difference of the dispersion relations is clearly visible from the critical exponents of $1/\beta$ 
which for $\Phi \not= n\Phi_0$ equal $2$ and $3$ respectively. In other words, in the small temperature regime, 
we find that the usual linear behaviour of the pure thermal noise as a function of the temperature (Johnson-Nyquist law) is changed
to a cubic behaviour in the presence of a magnetic flux for the Dirac dynamics. This is in sharp contrast with the change to a quadratic law
in the Schr\"odinger case. 
In principle, this fact could be used experimentally to determine whether a given junction  
is best described by the Schr\"odinger or the Dirac equation.

\section{Discussion and conclusion\label{conclu}}
We have presented a general method to construct an \textit{exact} total scattering matrix from given  scattering matrices localized on the vertices of a graph. The method is explicit and can be also applied in the presence of an ambient electromagnetic field. We found that, when the graph admits some symmetry, the properties of the total scattering matrix are entirely captured by its so-called fundamental eigenvalue, which we computed exactly.

The expression of the total scattering matrix can be used for the study of particle and energy transport on the graph. If the external edges of the graph are connected to heat bath with different temperatures and chemical potentials, the system is away from equilibrium and one can evaluate the correlation functions of the current flowing in the external edges. 

As an example, we considered the one-point and two-point current correlation functions in the case of a ring with $N$ external edges and a magnetic flux $\Phi$. Focusing first on the free Schr\"odinger case, we computed the total scattering matrix, and the quantum noise and their dependence on $\Phi$. When the \textit{local} scattering matrices are scale invariant and the chemical potentials vanish, the pure thermal noise behaves as $1/\beta^2$ at low temperatures for non-zero magnetic field, showing a deviation from the $1/\beta$ Johnson-Nyquist law. For non scale invariant local matrices (and/or non vanishing chemical potentials) one recovers the standard Johnson-Nyquist behavior, as shown on an example in the appendix \ref{appA} below.
Thus, the knowledge of local scattering matrices and the exact form of the total one are crucial to see this effect that cannot be observed by using effective total scattering matrices. Indeed, this effect on the thermal noise depends heavily on $k$-dependence of the \textit{total} scattering matrix. When the \textit{local} scattering matrices are scale invariant, this $k$-dependence comes only through the combination $kd$, where $d$ is the typical length of the ring: it can therefore be interpreted as a \textit{finite size} effect. This effect cannot be detected through other usual approaches, such as  conformal field theory, renormalization group techniques or bosonization.

We also performed the same analysis in the free massless Dirac case, obtaining a non-standard $1/\beta^3$ behaviour for the thermal noise. It is interesting to remark that the dynamics of the system is reflected in this non-standard behaviour.
\section*{Acknowledgments}
 M.M. is grateful to CNRS and Universit\'e de Savoie for financial support and hospitality during the early stage of the work. V.C. and E.R. acknowledge hospitality and financial support from INFN and Universit\`a di Pisa where this work was completed.

\appendix
\section{Non scale invariant local matrices\label{appA}}
So far, we have studied scale invariant local scattering matrices, which can be considered as critical points of a general scattering matrix. Indeed, the former are also fixed points of renormalization group flows.  
Suppose that instead of using $S^C$ to model the local scattering between the ring and the external leads, we use the following 
energy-dependent generalization of it
\bea
S^E(k)=\frac{1}{k+i\eta}
\left(
\begin{array}{ccc}
 (1-2 t )k-i \eta &  k\sqrt{2t(1-t)} &  k\sqrt{2t(1-t)} \\
  k\sqrt{2t(1-t)} & (t-1) k-i \eta & t k \\
  k\sqrt{2t(1-t)} & t k & (t-1) k-i \eta 
\end{array}
\right)
\,,
\eea
with $\eta\in\RR$ and $t\in(0,1)$. Note that $S^E(k)$ reproduces $S^C$ for $\eta=0$.
For $S^E(k)$, the fundamental eigenvalue reads
\bea
\lambda(k,\theta)=-\frac{t k (\cos (kd)-\cos (\theta ))+\eta \sin (kd)-i(t-1) k \sin (kd)}
{t k (\cos (kd)-\cos (\theta ))+\eta \sin (kd)+i(t-1) k \sin (kd)}\,.
\eea
Therefore, the small $k$ expansion now reads 
\bea\label{lambda-non-scale}
\lambda(k,\theta)=-1+\frac{2 i (t-1) d}{t(1-\cos \theta ) +d \eta }k+O\left(k^2\right)\,,
\eea
\textit{for all values of $\theta$} (assuming $\eta\neq0$). In other words, the expansion \eqref{lambda-non-scale} stands for both (\ref{exp:lambda0}) and (\ref{exp:lambda-theta}).

Consequently, when scale invariance is locally broken,  the magnetic flux has  no influence and the pure
thermal noise has the usual $1/\beta$ Johnson-Nyquist behaviour.


\begin{thebibliography}{99}

\bibitem{KF-a} C. L. Kane and M. P. A. Fisher, \textsl{Transport in a one-channel Luttinger liquid}, Phys. Rev. Lett. {\bf 68} (1992), 1220.

\bibitem{SS-a} I. Safi, H. J. Schulz, \textsl{Transport in an inhomogeneous interacting one-dimensional system}, Phys. Rev. {\bf B52} (1995), R17040.

\bibitem{NF-a} C. Nayak, M. P. A. Fisher, A. W. W. Ludwig and H. H. Lin, \textsl{Resonant multilead point-contact tunneling}, Phys. Rev. {\bf B59} (1999), 15694.

\bibitem{SD-a} I. Safi, P. Devillard, and T. Martin, Phys. Rev. Lett. 86, 4628 (2001).

\bibitem{MW-a} J.E. Moore and X.-G. Wen, Phys. Rev. B 66, 115305 (2002).

\bibitem{PPIL-a} K-V. Pham, F. Piechon, K-I Imura, P. Lederer, \textsl{Tomonaga-Luttinger liquid with reservoirs in a multi-terminal geometry}, Phys. Rev. {\bf B68} (2003), 205110
and \texttt{arXiv:cond-mat/0207294}.

\bibitem{KD-a} K. Kazymyrenko and B. Doucot, \textsl{Regular networks of Luttinger liquids}, Phys. Rev. {\bf B71} (2005), 075110 and \texttt{arXiv:cond-mat/0407268}.

\bibitem{SRS-a} S. Das, S. Rao, D. Sen, \textsl{Inter-edge interactions and novel fixed points at a junction of quantum Hall line junctions}, Phys. Rev. {\bf B74} (2006), 
045322 and \texttt{arXiv:cond-mat/0511097};\\
  S. Das and S. Rao, \textsl{Duality between normal 
and superconducting junctions of multiple quantum wires }, Phys. Rev. {\bf B78} (2008), 205421 and \texttt{arXiv:0807.0804};\\
 A. Agarwal, S. Das, S. Rao and D. Sen, 
\textsl{Enhancement of tunneling density of states at a junction of three Luttinger liquid wires}, Phys. Rev. Lett. {\bf 103} (2009), 026401 and 
\texttt{arXiv:0810.3513}; Erratum, Phys. Rev. Lett. {\bf 103}, 079903 (2009);\\
A. Soori, D. Sen, \textsl{Conductance of Tomonaga-Luttinger liquid wires and junctions with resistances}, Europhys. Lett. {\bf 93} (2011), 57007 and \texttt{arXiv:1011.5058}.

\bibitem{Mintchevetal} B. Bellazzini and M. Mintchev, \textsl{Quantum Fields on Star Graphs}, J. Phys. {\bf A39} (2006), 11101 and \texttt{arXiv:hep-th/0605036};\\
B. Bellazzini, M. Burrello, M. Mintchev and P. Sorba, \textsl{Quantum Field Theory on Star Graphs}, Proc. Symp. Pure Math.\textbf{77} (2008) 639 and \texttt{arXiv:0801.2852};\\
 B. Bellazzini, P. Calabrese and M. Mintchev, \textsl{Junctions of anyonic Luttinger wires}, Phys. Rev. {\bf B79} (2009), 085122 and \texttt{arXiv:0808.2719};\\
 B. Bellazzini, M. Mintchev and P. Sorba, \textsl{Quantum Fields on Star Graphs with Bound States at the Vertex}, J. Math. Phys. {\bf 51} (2010), 032302
and \texttt{arXiv:0810.3101}; \\
 B. Bellazzini, M. Mintchev and P. Sorba, \textsl{Off-critical Luttinger Junctions}, Phys. Rev. {\bf B82} (2010), 195113 and \texttt{arXiv:1002.0206}.

\bibitem{Yi-b} H. Yi, \textsl{Resonant tunneling and the multichannel Kondo problem: Quantum Brownian motion description}, Phys. Rev. {\bf B65} (2002), 195101 and 
\texttt{arXiv:cond-mat/9912452}.

\bibitem{LRS-b} S. Lal, S. Rao, and D. Sen, \textsl{Junction of several weakly interacting quantum wires: a renormalization group study}, Phys. Rev. {\bf B66} (2002), 165327
and \texttt{arXiv:cond-mat/0206259}.

\bibitem{EMAB-b} T. Enss, V. Meden, S. Andergassen, X. Barnabe-Theriault, W. Metzner, K. Sch\"onhammer, \textsl{Correlation effects on resonant tunneling in one-dimensional quantum wires},
Phys. Rev. \textsl{B71}, 155401 (2005) and \texttt{arXiv:cond-mat/0403655}; \\
X. Barnabe-Theriault, A Sedeki, V. Meden, K. Sch\"onhammer,
\textsl{A junction of three quantum wires: restoring time-reversal symmetry by interaction},
Phys. Rev. Lett. {\bf 94} (2005), 136405 and \texttt{arXiv:cond-mat/0411612}; \\
X. Barnabe-Theriault, A. Sedeki, V. Meden,
K. Sch\"onhammer, \textsl{Junctions of one-dimensional quantum wires - correlation effects in transport}, Phys. Rev. {\bf B71} (2005), 205327 and \texttt{arXiv:cond-mat/0501742}.

\bibitem{DRS-b} S. Das, S. Rao, and A. Saha, \textsl{Renormalization group study of transport through a superconducting junction of 
multiple one-dimensional quantum wires}, Phys. Rev. {\bf B77} (2008), 155418 and \texttt{ arXiv:0711.1324};

\bibitem{COA-c} C. Chamon, M. Oshikawa, and I. Affleck, \textsl{Junctions of three quantum wires and the dissipative Hofstadter model}, Phys. Rev. Lett. {\bf 91} (2003), 206403 
and \texttt{arXiv:cond-mat/0305121}; \\
M. Oshikawa, C. Chamon, and I. Affleck, \textsl{Junctions of three quantum wires}, J. Stat. Mech. {\bf 602} (2006), P02008 and \texttt{ arXiv:cond-mat/0509675}.

\bibitem{Affetal} A. Rahmani, C-Y. Hou, A. Feiguin, M. Oshikawa, C. Chamon, I. Affleck 
, \textsl{General method for calculating the universal conductance of strongly correlated junctions of multiple quantum wires},
Phys. Rev. {\bf B85} (2012), 045120 and \texttt{arXiv:1108.4418};\\
  C-Y. Hou, A. Rahmani, A. E. Feiguin, C. Chamon, \textsl{Junctions of multiple quantum wires with different Luttinger parameters}, 
Phys. Rev. {\bf B86} (2012), 075451 and \texttt{ arXiv:1205.2125}. 

\bibitem{CTE-d} S. Chen, B. Trauzettel, and R. Egger, \textsl{Landauer-type transport theory for interacting quantum wires: Application to carbon nanotube Y junctions},
 Phys. Rev. Lett. {\bf 89} (2002), 226404 and \texttt{arXiv:cond-mat/0207235}.

\bibitem{la-57}
R. Landauer, \textsl{Spatial variation of currents and fields due to localized scatterers in metallic conduction},
IBM J. Res. Dev. {\bf 1} (1957), 233;\\
---, \textsl{Electrical resistance of disordered one-dimensional lattices},
Philos. Mag. {\bf 21} (1970), 863.

\bibitem{bu-86}
M. B\"uttiker, 
\textsl{Four-Terminal Phase-Coherent Conductance}, Phys. Rev. Lett. {\bf 57} (1986), 1761; \\
---, \textsl{Symmetry of electrical conduction}, IBM J. Res. Dev. {\bf 32} (1988), 317. 

\bibitem{CR} V. Caudrelier, E. Ragoucy, \textsl{Direct computation of scattering matrices for general quantum graphs}, Nucl. Phys. {\bf B828} (2010), 515
and \texttt{arXiv:0907.5359}. 

\bibitem{KS} V. Kostrykin, R. Schrader, \textsl{Kirchhoff's Rule for Quantum Wires}, J. Phys. {\bf A32} (1999), 595 and 
\texttt{arXiv:math-ph/9806013}.

\bibitem{biy-84} M. B\"uttiker, Y. Imry and M. Ya. Azbel, \textsl{Quantum oscillations in one-dimensional normal-metal rings},
Phys. Rev.  A {\bf 30} (1984), 1982.

\bibitem{CRM}  V. Caudrelier, M. Mintchev, E. Ragoucy, \textsl{Quantum Wire Network with Magnetic Flux}, Phys. Lett. {\bf A377} (2013), 1788
and \texttt{arXiv:1202.4270}.

\bibitem{MRS} M. Mintchev, E. Ragoucy, P. Sorba, \textsl{Reflection-Transmission Algebras}, J. Phys. {\bf A36} (2003), 10407
and \texttt{arXiv:hep-th/0303187}.

\bibitem{KS1} V. Kostrykin, R. Schrader, \textsl{Quantum wires with magnetic fluxes}, Comm. Math. Phys. {\bf 237} (2003), 161 and 
\texttt{arXiv:math-ph/0212001}.

\bibitem{KS0} 
V.~Kostrykin and R.~Schrader, \textsl{The Generalized Star Product and the Factorization of Scattering Matrices on Graphs }, 
J. Math. Phys. {\bf 42} (2001) 1563.

\bibitem{KoSm} 
T. Kottos and U. Smilansky, \textsl{Chaotic Scattering on Graphs}, Phys. Rev. Lett. {\bf 85} (2000) 968. 


\bibitem{Mintchev:2011mx}
M.~Mintchev, \textsl{Non-equilibrium Steady States of Quantum Systems on Star Graphs}, J. Phys. A  {\bf 44} (2011), 415201 and \texttt{arXiv:1106.5871}.

\bibitem{bb-00} 
Ya.~M.~Blanter and M.~B\"uttiker, 
\textsl{Shot Noise in Mesoscopic Conductors}, Phys. Rep. {\bf 336} (2000), 1. 

\bibitem{BMS}  Brando Bellazzini, Mihail Mintchev, Paul Sorba , \textsl{Quantum wire junctions breaking time reversal invariance}, Phys. Rev. {\bf B80} (2009) 
, 245441 and  \texttt{arXiv:0907.4221}.

\end{thebibliography}
\end{document}